\documentclass{jfm}
\usepackage{graphicx}
\usepackage{epstopdf, epsfig}
\usepackage{siunitx}
\usepackage[dvipsnames]{xcolor}
\usepackage{amsmath}
\usepackage{cleveref}
\usepackage{wasysym}
\usepackage{amssymb}

\definecolor{skyblue}{rgb}{0.00, 0.45, 0.74} 

\crefname{section}{§}{§§}
\Crefname{section}{§}{§§}

\shorttitle{Rapidly pitching plates in decelerating motion}
\shortauthor{Dibya R. Adhikari and Samik Bhattacharya}

\title{Rapidly pitching plates in decelerating motion near the ground}

\author{Dibya R. Adhikari\aff{1}
  \corresp{\email{dibyarajadhikari@knights.ucf.edu}},
 \and Samik Bhattacharya\aff{1}}

\affiliation{\aff{1}Department of Aerospace and Mechanical Engineering, University of
Central Florida, 4000 Central Florida Blvd, Orlando, FL 32816, USA}

\begin{document}

\maketitle

\begin{abstract}
Birds employ rapid pitch-up motions for different purposes: perching birds use this motion to decelerate and come to a complete stop while hunting birds, like bald eagles, employ it to catch prey and swiftly fly away. Motivated by these observations, our study investigates how natural flyers accomplish diverse flying objectives by rapidly pitching their wings during deceleration. We conducted experimental and analytical investigations focusing on rapidly pitching plates during deceleration in close proximity to the ground to explore the impact of ground proximity on unsteady dynamics. Initially, we executed simultaneous deceleration and pitch-up motion close to the ground. Experimental results demonstrate that as the pitching wing approaches the ground, the instantaneous lift increases while the initial peak drag force remains relatively unchanged. Our analytical model conforms this trend, predicting an increase in lift force as the wing approaches the ground, indicating enhanced added mass and circulatory lift force due to the ground effect. Next, we examined asynchronous motion cases, where rapid pitching motions were initiated at different stages of deceleration. The results reveal that when the wing pitch is synchronized with the start of deceleration, larger counter-rotating vortices form early in the maneuver. These vortices generate stronger dipole jets that orient backward in the later stages of the maneuver after impinging with the ground surface, which hunting birds recover to accelerate after catching prey. Conversely, when the wing pitch is delayed, smaller vortices form, but their formation is postponed until late in the maneuver. This delayed vortex formation generates beneficial unsteady forces late in the maneuver that facilitates a smooth landing or perching. Thus, through strategic tuning of rapid pitch-up motion with deceleration, natural flyers, such as birds, achieve diverse flying objectives. 
\end{abstract}

\begin{keywords}
Perching wing, hunting, rapid area change, ground effect, dipole jet, 
\end{keywords}

\section{Introduction}
Birds have been observed to pitch their wings for a variety of purposes: such as perching birds that rapidly pitch their wings upward to decelerate to a complete stop while landing \citep{carruthers_2007, carruthers_2010, berg_2010, Provini_2014}, and hunting birds, which also use the same motion to slow down to catch a fish out of water before flying away \citep{reimann_1938, todd_1982, venable_1996, stalmaster_1997, gerrard_2014, sorensen_2015, collard_2021}. A review of existing research revealed that there are several basic aerodynamic questions related to pitching plates in deceleration that have not been sufficiently addressed in the current literature. These include: (1) how ground proximity affects the unsteady dynamics of the perching maneuver, and (2) how birds achieve different flying objectives by rapidly pitching their wings during deceleration. In this paper, we aim to investigate these questions using experimental and analytical approaches. 

Reducing the distance between the wing and the ground increases the lift-to-drag ratio due to the ground effect \citep{zerihan_2000, luo_2012}. The use of the ground effect has been observed in many natural flyers and swimmers \citep{baudinette_1974, hainsworth_1988, withers_1977, blake_1979, webb_1993, saffman_1967, park_2010}, which have evolved to take advantage of the ground effect to enhance their performance. This increase in operational efficiency has inspired the design of wings in ground effect aircraft \citep{rozhdestvensky_2006}. 

Numerous efforts have been made in recent decades to understand the ground effect experienced by an oscillating airfoil close to the ground \citep{fernandez_2015, mivehchi_2016, zhang_2017}. Studies have revealed that the effects of ground proximity on unsteady aerodynamics can vary depending on the type of wing kinematics. For instance, in the study by \citet{quinn_2014} on a pitching airfoil, they found that when the airfoil is close to the ground, it experiences increased lift force that pushes the airfoil away from the ground. They also observed that pitching near the ground generates a vortex pair instead of a vortex street, increasing the average thrust force. In another study, \citet{deepthi_2021} showed that an inclined flapping wing in ground effect experiences a higher lift at a stroke plane angle of $45^\circ$, but not at other angles. Therefore, to fully quantify the performance of perching birds, further exploration of the ground effect experienced by a rapid pitching plate during deceleration is necessary. 

Rapid pitching causes a quick change in the surface area of the wing facing the airflow, which can significantly impact the airflow over the wing. This rapid change in the wing's surface area can have fascinating prospects for flow control, as the added mass changes with the change in the surface area, which can play a significant role in varying the dynamic forces on the wing. \citet{saffman_1967} showed that a body could propel itself by deforming its surface area through added mass recovery. \citet{childress_2006} conducted an experimental study on a flexible body in oscillating air. They concluded that exposing the variable frontal area to the airflow due to wing flapping leads to changes in the added mass, resulting in stable hovering. Rapid area change can lead to boundary layer separation and the shedding of the vortices on a vanishing body \citep{wibawa_2012}. \citet{weymouth_2013} concluded that in addition to added mass recovery, the deforming body relies on flow separation elimination to achieve ultra-fast escape. However, our understanding of rapid area change is limited to decreasing surface area in acceleration, and the study of how the flow behaves over an increased surface area during deceleration, which is commonly experienced by birds that perch and hunt, has yet to be extensively studied.

Recently, \citet{polet_2015} conducted an experimental and numerical study on the unsteady aerodynamics of a two-dimensional NACA0012 airfoil undergoing simultaneous pitch-up and decelerating motion. They found that the significant lift and drag force on a wing during perching is mainly caused by the added mass effect and the formation of strong vortices at the leading and trailing edge of the wing. \citet{jardin_2019} also performed a numerical study on a perching airfoil and concluded that a minimum kinetic energy could be achieved on the airfoil at the end of the perching maneuver at a higher pitch rate or the lift and drag force on the airfoil can be enhanced by increasing the pitch rate. \citet{Dibya_2022} also studied the unsteady dynamics of a finite wing undergoing a rapid pitch-up motion while decelerating and descending close to the ground. They showed that a perching wing could generate higher forces by using a combination of pitching and heaving motions during deceleration. However, in these studies, although the wing generated a higher drag force by increasing the pitch rate, which is appropriate for decelerating rapidly to a complete stop, the perching wing also generated a higher lift force. This higher lift force causes the wing to rise in altitude, which may not be desirable for perching at the initial perching location or altitude. Thus, more research is needed to understand the mechanics involved when the wing pitches rapidly during deceleration.

Studies on rapid area change have also investigated the effect of varying the frontal area against the incoming airflow. \citet{spagnolie_2009} found through a numerical simulation that by controlling the phase difference between the shape change and background flow of an oscillating flow, a shape-changing body can generate vortex structures that induce a downward moving dipole jet below the body. The resulting jet of fluid enabled the body to hover or ascend vertically. Similarly, \citet{weymouth_2013} showed that, by deforming quickly, a rapidly shape-changing body could eliminate the flow separation from its surface. This reduces drag and increases thrust force, which is beneficial for escape maneuvers. While these studies provide insight into the timing of the shape change with the airflow and its impact on the added mass and vortex evolution, they were limited to oscillating or accelerating flows. Therefore, this paper focuses on varying the timing of rapid pitch-up motion relative to decelerating flow and assesses the impact of such motion on the perching and hunting birds' performance.

In this paper, we consider two scenarios of rapid pitch-up motion of plates in decelerating motion near the ground to better understand the aerodynamic mechanism used by perching and hunting birds. In the first scenario, we perform a synchronous motion where the wing pitches up while decelerating from steady velocity to a complete stop, with the same motion duration for both motions. We execute synchronous motion at different ground heights to understand the ground effect on perching plates. In the second scenario, we create an asynchronous motion where the deceleration time is longer than the pitch time of the plate, allowing the execution of the pitch-up motion at various stages of deceleration. By comparing the evolution of unsteady forces and flow field while varying the start of pitch-up motion during deceleration, the current study aims to gain new insights into the aerodynamic mechanisms natural flyers use and how these mechanisms can be replicated in the design of next-generation flying vehicles and aircraft. 

\begin{figure}
  \centerline{\includegraphics[scale=0.5]{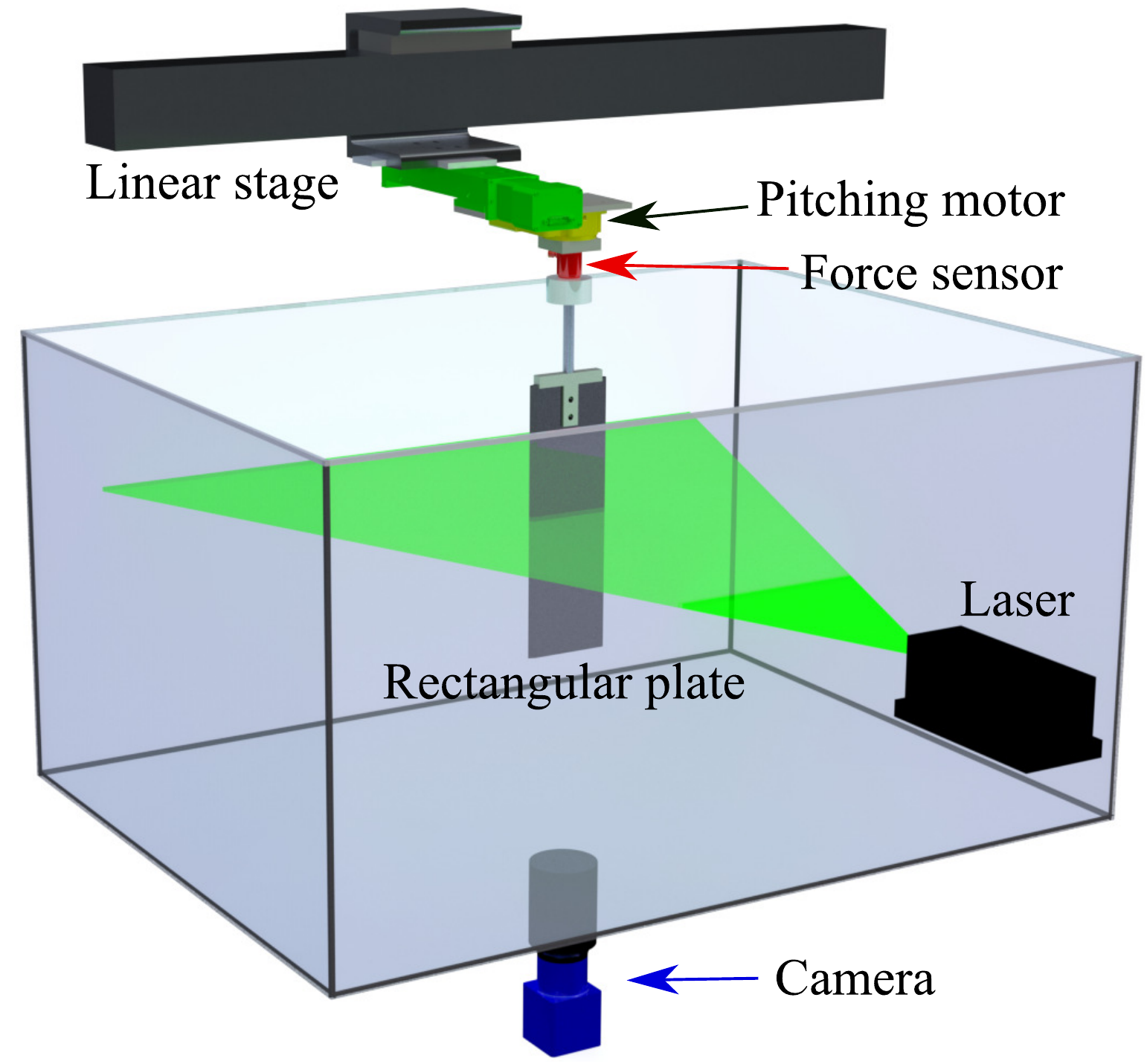}}
  \caption{Schematic diagram of the experimental set-up illustrating the rectangular plate, with details on the placement of force sensor, camera, and laser.}
\label{fig: Experimental set-up}
\end{figure}

\section{Methodology}\label{sec:methods}
\subsection{Experimental setup}
The schematic diagram of the experimental setup is shown in figure \ref{fig: Experimental set-up}. The tests were conducted in a water-filled towing tank with a free surface measuring \SI{0.9}{m} in length, {\SI{0.45}{m}} in width, and {\SI{0.4}{m}} in height. The wing model was mounted on a liner stage powered by a servo motor (FSL120, FUYU Inc., China), which moved it along the length of the towing tank. The wing model's deceleration was prescribed by gradually slowing down the linear stage. A stepper-driven linear stage (LSQ150B-T3, Zaber Tech. Inc., Canada) connected orthogonally to the servo-driven stage moved the wing model towards the solid boundary, which acted as a ground in this study. A rotary stage powered by a stepper motor (RSW60A-T3, Zaber Tech. Inc., Canada) executed the rapid pitch-up motion around the mid-chord of the wing model. A force sensor was installed on the setup below the pitching motor and was connected to the wing model via a \textcolor{red}{\SI{0.09}{m}} thick cylindrical rod. A pulse generator (9400 series, Quantum Composers Inc., USA) sent a trigger pulse signal to synchronize the deceleration motion, the start of the pitching motor, the force sensor, and the camera. The wing model was submerged vertically in the tank, with the wingtip positioned \SI{0.2}{m} from the bottom. 

\begin{table}
  \begin{center}
\def~{\hphantom{0}}
  \begin{tabular}{lccccc}
      \text{Nomenclature}  &~$\dot{U}~\SI{}{(m/s^2)}$   &   $\dot{\alpha}~\SI{}{(rad/s)}$ & $\Xi$ & \text{Synchronous} & \text{Asynchronous ($t^*_{os}$)}\\[3pt]
       ~~~~~C1   & 0.0351 & ~0.550~ & 0.2 & \checkmark & $\times$\\
       ~~~~C2\_0   & 0.0234 & ~0.550~ & 0.3 & $\times$ & \checkmark $(0.00)$\\
       ~~~~C2\_45   & 0.0234 & ~0.550~ & 0.3 & $\times$ & \checkmark $(0.25)$\\
       ~~~~C2\_90   & 0.0234 & ~0.550~ & 0.3 & $\times$ & \checkmark $(0.50)$\\
       ~~~~~C3   & 0.0877 & ~1.377~ & 0.4 & \checkmark & $\times$\\
       ~~~~C4\_0    & 0.0585 & ~1.377~ & 0.6 & $\times$ & \checkmark $(0.00)$\\
       ~~~~C4\_45   & 0.0585 & ~1.377~ & 0.6 & $\times$ & \checkmark $(0.25)$\\
       ~~~~C4\_90   & 0.0585 & ~1.377~ & 0.6 & $\times$ & \checkmark $(0.50)$\\
       ~~~~~C5   & 0.1111 & ~1.745~ & 0.6 & \checkmark & $\times$\\
       ~~~~C6\_0    & 0.0741 & ~1.745~ & 0.9 & $\times$ & \checkmark ~$(0.00)$\\
       ~~~~C6\_45   & 0.0741 & ~1.745~ & 0.9 & $\times$ & \checkmark $(0.25)$\\
       ~~~~C6\_90   & 0.0741 & ~1.745~ & 0.9 & $\times$ & \checkmark $(0.50)$\\
  \end{tabular}
  \caption{Summary of the kinematic parameters}
  \label{tab: Kinematic parameters}
  \end{center}
\end{table}

\begin{figure}
  \centerline{\includegraphics[width=1.0\textwidth]{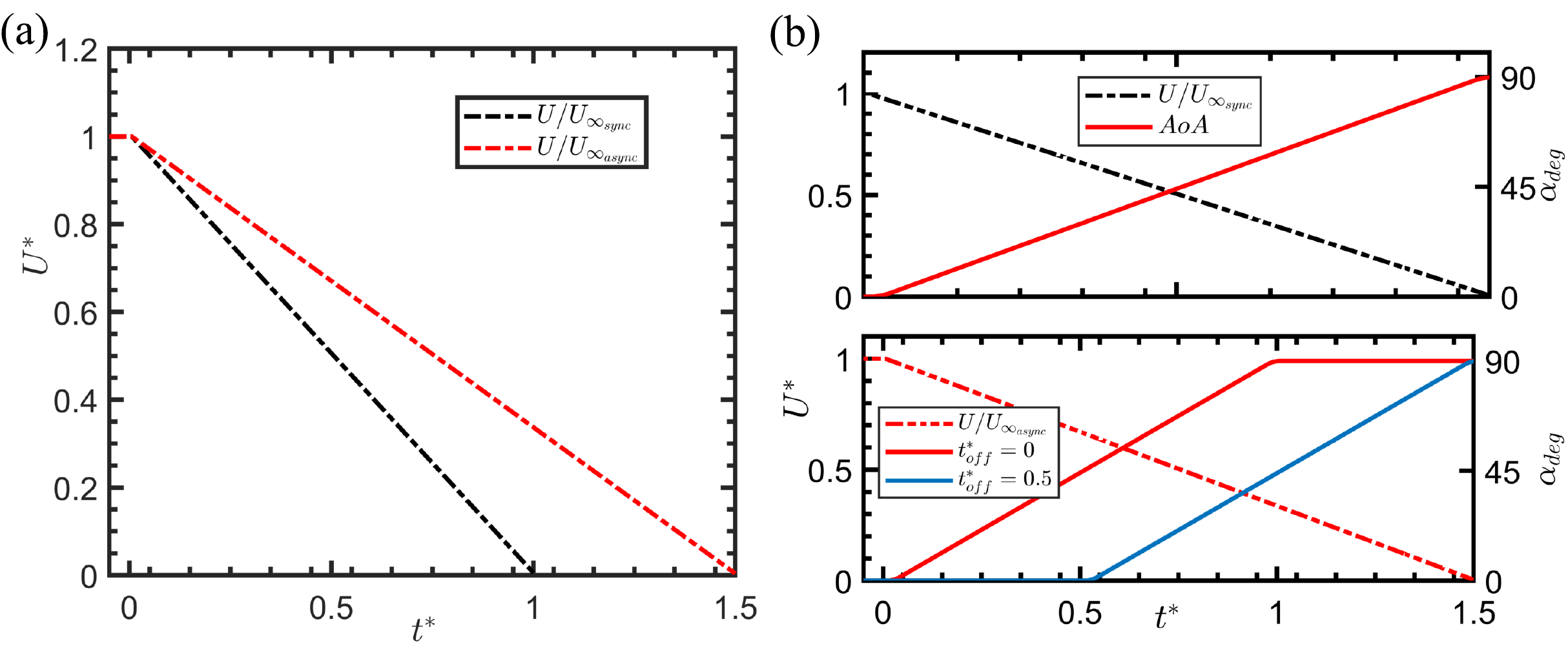}}
  \caption{(a) Comparison of the variation of non-dimensional velocity $U^*$ between synchronous and asynchronous motions. Variation of velocity and angle of attack for two perching scenarios: (b) synchronous motion; and (c) asynchronous motion. Comparisons shown are represented as a function of non-dimensional time $t^* = t/t_{p}$, where $t_{p}$ is the time period of pitch-up motion. Here, the decelerating velocity is scaled by the steady state velocity, $U_\infty$, while the total change in the angle, which is $90^0$, scales the angle of attack during pitch-up motion. The ratio of the time period of deceleration between synchronous, $t_{{d}_s}$, and asynchronous motion, $t_{{d}_{as}}$, is $t_r = \frac{t_{{d}_s}}{t_{{d}_{as}}} = \frac{1}{1.5}$. For the synchronous case, $t_{{d}_s} = t_{p_{s}}$, whereas for asynchronous case, $t_{{d}_{as}} = 1.5* t_{p_{as}}$.}
\label{fig: Description of the kinematics}
\end{figure}

\subsection{Wing model and problem description}
We used a finite rectangular wing planform with a chord length ($c$) of \SI{0.05}{m} and a planform area of \SI{0.0075}{m^2} as a perching wing model. The aspect ratio ($AR$) of the wing was \SI{3}{} and was fabricated from \SI{6}{mm} thick flat aluminum plate. The wing's leading edge (LE) was rounded, and the trailing edge (TE) was sharpened to meet the Kutta condition, ensuring the flow smoothly leaves the TE. 

 To simulate the perching maneuvers, two scenarios were considered: synchronous motion and asynchronous motion. In both scenarios, the wing model was initially oriented at an angle of attack $\alpha_0 = 0^0$ and then rapidly pitched up to $\alpha = 90^0$ while undergoing deceleration. 

In synchronous motion, the wing decelerated from steady velocity $U_{\infty}$ of \SI{0.1}{\metre\per\second} to a complete stop while pitching up, with both motions having the same motion duration ($t^*_{decn} = t^*_{pitch}$; see Figure \ref{fig: Description of the kinematics} (b)). This means that the start and end of the deceleration and rapid pitch-up motions are synchronized. The pitch-up motion causes a rapid increase in the frontal area of the wing facing the flow. This increase, combined with simultaneous deceleration, was quantified using the shape change number:

\begin{equation}
  \Xi=\frac{V}{\triangle U}
  \label{Helm}
\end{equation}

where $V$ is the frontal area speed of the airfoil and $\triangle U$ is the change in the translation speed of the wing during deceleration. We executed three shape change numbers ($\Xi$ = 0.2, 0.4, $\&$ 0.6) at ten non-dimensional ground heights ranging from $h^*$ = $\frac{h}{c}$ = $1.5-0.04$. We refer to $h^*$ = 1.5 as far from the ground case and $h^*$ = 0.04 as close to the ground case.  

 In asynchronous motion, the deceleration time was extended compared to synchronous (see Figure \ref{fig: Description of the kinematics} (a)), while keeping the pitching time constant. This results in time offsets between the two motions, with the deceleration time longer than the time to pitch, i.e. $t^*_{decn_{as}}$ = $1.5*t^*_{pitch_{as}}$. This time offset allows the pitch-up motion to be executed at various stages of the deceleration. As a result, for the same pitch rate, the change in velocity when the wing completes the pitch-up motion is $\triangle U = U_{\infty}*\frac{t^*_{pitch}}{t^*_{decn}} = U_{\infty}*\frac{1}{1.5}$, resulting in a higher shape change number $\Xi$ for asynchronous motion compared to synchronous case, i.e. $\Xi_{asyn} = 1.5*\Xi_{syn}$. For each $\Xi$, we considered three starting time offsets ($t^*_{os}$ = 0, 0.25, and 0.5) between the decelerating and pitch-up motion. When $t^*_{os} = 0$, the start of the deceleration and pitch-up motions are in sync, but the pitch-up motion ends before the wing decelerates to a complete stop. With $t^*_{os} = 0.5$, the start of the pitch-up motion lags the start of the deceleration motion, but the end of the pitch-up motion and deceleration motion is synchronized. Each asynchronous motion case was executed at three non-dimensional ground distance $h^*$ = 1.5, 0.25, and 0.04. 

 The steady-state velocity of the wing model was ($U_{\infty}$) of \SI{0.1}{\metre\per\second}. The Reynolds number of the perching wing model, based on $U_{\infty}$ and $c = \SI{0.05}{m}$, was \textcolor{red}{$Re = 6,500$}. Table \ref{tab: Kinematic parameters} summarizes the kinematic parameters used in this experiment.


\subsection{Measurement of Instantaneous forces}
We measured the instantaneous forces acting on the wing using a six-axis force and torque sensor (NANO 17, ATI Inc., USA) connected to a 16-bit DAQ device (NI-USB-6211, National Instrument, USA). The force-sensor data was collected at a sampling rate of \SI{5}{\kilo\hertz} and averaged over five test runs. The combined wing motion produced an oscillatory frequency of approximately \SI{4}{Hz} on the force-sensor data. We filtered the force-sensor data with a Butterworth low pass filter with a cutoff frequency of \SI{3}{Hz} to remove this vibration, while retaining most of the fluid force oscillatory peaks. Subsequently, we smoothed the data using the moving average of 20 points. We found that the uncertainty in the force data is found to be around $6\%$ at the peak and less than $3\%$ for the smaller magnitude of the forces.

To account for the inertial forces and the weight of the wing assembly, we performed both dynamic and static tare experiments. In dynamic tare, we conducted tare experiments in the air using the same kinematics as in the water. We observed that the lift force in water was approximately 11 times higher than in the air. Since the apparent mass of the water accelerated along the model was approximately 10 times higher than the mass of the model and the force balance, we neglected the dynamic tare in the air, following the approach proposed by ~\citet{barlow_1999} and \citet{granlund_2013}. In static tare, we measured the data in still water every $3^{\circ}$ of the pitch angle up to the maximum pitch angle. The wing model produced negligible static tare, so its contribution was not considered in our analysis.

\subsection{PIV measurements}
We measured the velocity field at the 50\% wing span using Planar particle image velocimetry (PIV). To seed the water tank, we used neutrally buoyant, \SI{10}{\micro\meter} diameter silver-coated hollow glass spheres (Conductophill, Potter Inc., USA). The laser sheet for illuminating the plane of interrogation was generated by a \SI{2}{mm} diameter beam from a continuous-wave green laser \textcolor{red}{(DPSS-DMPV-532-2, Egorov Scientific, USA)}, which was expanded into a \SI{2}{mm} thick laser sheet by using two cylindrical lenses. We recorded images of the illuminated plane by a \textcolor{red}{AOS camera (J-Pri, AOS Tech. AG, Switzerland)} at a frame rate of \SI{200}{\hertz} and a resolution of 2560 x 1920 pixels.  The field of view was \SI{0.25}{\meter} x \SI{0.18}{\meter}. The images were processed in PIVLab, a MATLAB-based software. We used a multi-pass iterative algorithm with a window size of 64 x 64 pixels in the first pass and 32 x 32 pixels in the second pass, with a $50\%$ overlap between successive windows. Finally, we phase-averaged the PIV data over five runs.

\section{Results and Discussion}\label{sec:Results_Discussion}
In \cref{sec: Synchronous motion results} and \cref{sec: Asynchronous motion results}, we present the results separately for synchronous and asynchronous motion, where synchronous motion refers to cases where the start and end of deceleration and pitch-up motions are in synchrony, while asynchronous motion refers to the cases where the two motions are not in synchrony. We then discuss the dipole jet induced due to counter-rotating vortices in \cref{sec: Discussion on Dipole Jet}. Next, in \cref{sec: Scaling laws results}, we focus on the scaling laws for perching maneuvers. Finally, in \cref{sec: Analytical model and results}, we compare instantaneous forces between the experimental and analytical model results.

\begin{figure}
\centerline{\includegraphics[width=1.0\textwidth]{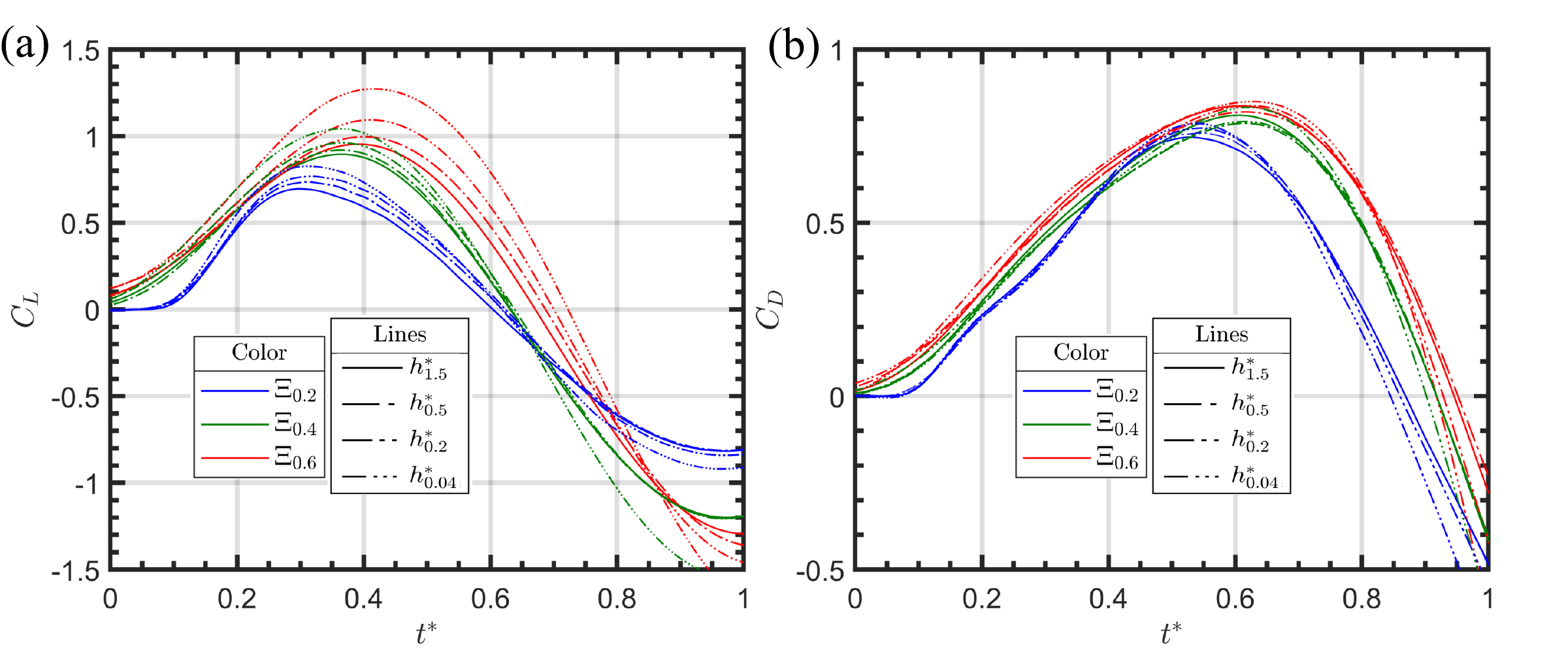}}
  \caption{Comparisons of (a) lift and (b) drag coefficient on the rectangular wing for $\Xi$ = 0.2, 0.4, and 0.6. Each $\Xi$ are presented at four non-dimensional ground height, $h^*$.}
\label{fig: Synchronous motion lift and drag coefficient}
\end{figure}

\subsection{Synchronous motion: Unsteady forces and flow-field}\label{sec: Synchronous motion results}
Figure \ref{fig: Synchronous motion lift and drag coefficient} displays the evolution of unsteady lift and drag forces during synchronous motion at three different shape change numbers $\Xi =$ 0.2, 0.4, and 0.6. To provide a comprehensive analysis of the ground effect, we executed each $\Xi$ at ten different non-dimensional ground heights ranging from $h^* = 1.5-0.04$. For a clear and concise representation of the plot, we provided unsteady forces at four different values of $h^*$. The deceleration and pitch-up motion were both initiated at $t^* = 0$, and the unsteady forces were measured during perching. 

Figure \ref{fig: Synchronous motion lift and drag coefficient}(a) shows that the execution of simultaneous deceleration and pitch-up motion results in a steep rise in the lift coefficient and attains the peak value after a certain time instant. This initial rise in the lift coefficient is mainly due to the combined effect of non-circulatory and circulatory force. The plot indicates that the peak lift force coefficient increases with increasing $\Xi$, which is consistent with the results of \citet{polet_2015} and \citet{jardin_2019}. The peak lift coefficient increases by approximately \textcolor{red}{50\%} as $\Xi$ increases from 0.2 to 0.6. After this initial peak force, the wing experiences a decline in the lift coefficient. This decay in the lift is related to the detachment of the leading edge vortex (LEV) from the wing leading edge (LE), leading to the stalled flow at the latter stage of the motion. This decline in lift coefficient can also be related to the decrease in the non-circulatory force due to the deceleration of the wing. From figure \ref{fig: Synchronous motion lift and drag coefficient} (a), it is observed that the lift force for $\Xi = 0.6$ starts to decay at a later stage of the motion compared to that of $\Xi = 0.2$.  

Figure \ref{fig: Synchronous motion lift and drag coefficient} (a) also illustrates the effect of ground proximity on the instantaneous lift coefficient of the perching plates at various non-dimensional ground heights ranging from $1.5 \leq h^* \leq 0.04$. As $h^*$ decreases, the initial rise in the lift force increases consistently for each $\Xi$. For $\Xi = 0.2$, the initial peak lift force rises by approximately $19\%$, whereas for $\Xi = 0.6$, this rise is approximately $38\%$. Although the initial peak force increases with ground proximity, the perching plate also experiences an increase in the trough force at the end of the maneuver as the wing approaches the ground. However, for the majority of the perching maneuver, the instantaneous life force increases when the wing is close to the ground, which can be beneficial during this highly unsteady maneuver. 

\begin{figure}
\centerline{\includegraphics[width=1.0\textwidth]{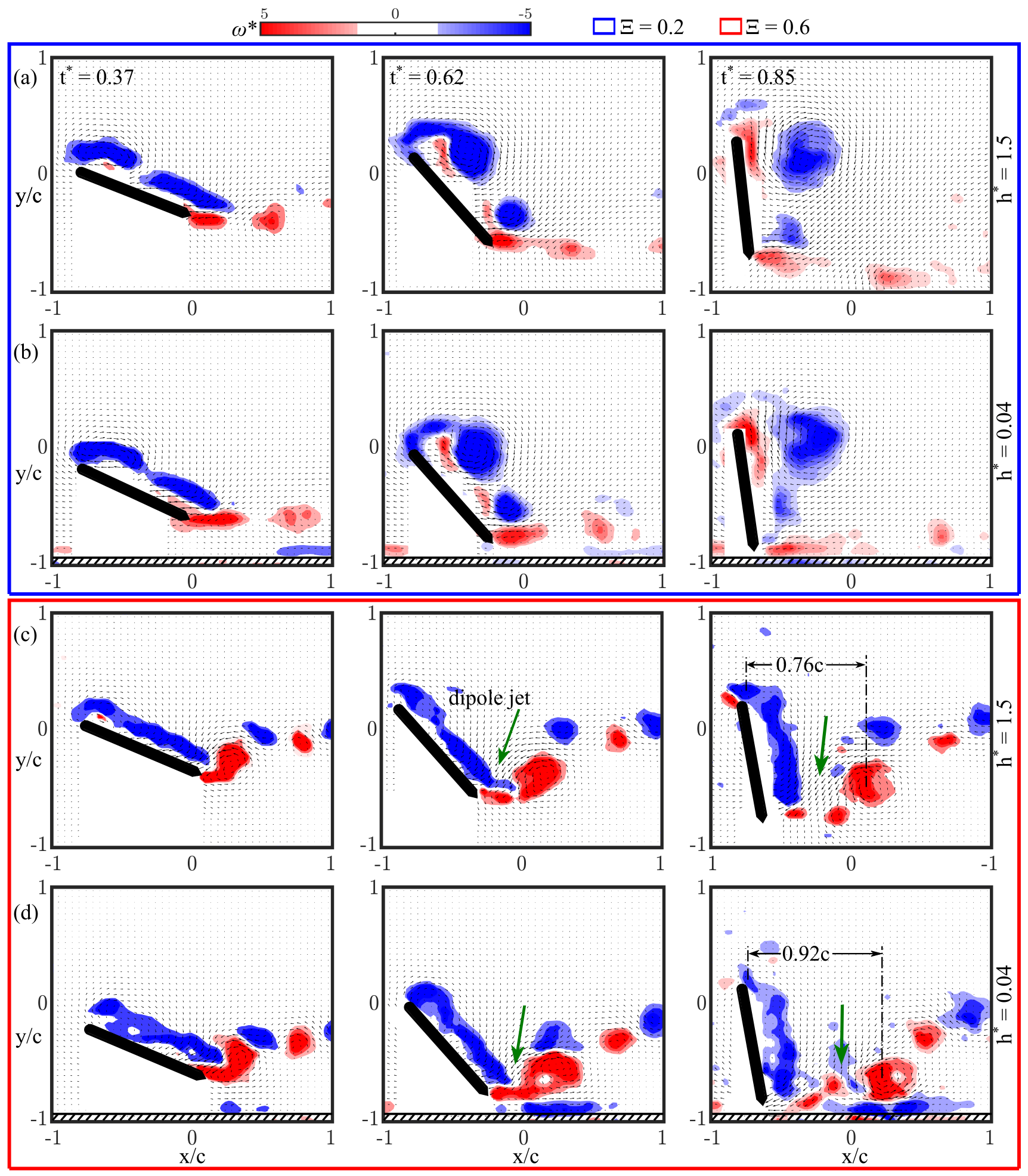}}
  \caption{Contours of the normalized vorticity field, $\omega^*$, for synchronous motion at the $50\%$ of the wing span at three time-steps, $t^*$ = 0.37, 0.62, \& 0.85: $\Xi$ = 0.2 at (a) $h^*$ = 1.5; (b) $h^*$ = 0.04. $\Xi$ = 0.6 at (c) $h^*$ = 1.5; (d) $h^*$ = 0.04.}
\label{fig: Synchronous vorticity field}
\end{figure}

The evolution of the drag coefficient on the perching plate is shown in figure \ref{fig: Synchronous motion lift and drag coefficient} (b). It is observed that higher values of $\Xi$ lead to a larger instantaneous drag coefficient, with the peak drag coefficient increasing by approximately $10\%$ when $\Xi$ is changed from 0.2 to 0.6. Interestingly, the time instant of the initial peak drag force shifts from \textcolor{red}{$t^* = 0.45$ for $\Xi = 0.2$ to $t^* = 0.52$ for $\Xi = 0.6$}. This trend is similar to that observed for the lift coefficient, which suggests that the peak and decay of the forces occur at a higher angle of attack for higher values of $\Xi$. This phenomenon has also been observed by \citet{kleinheerenbrink_2022}, who concluded that perching birds pitch faster to minimize the stall distance.

In contrast to the lift, the evolution of the drag coefficient is not significantly affected by ground proximity. For each $\Xi$, varying $h^*$ leads to a negligible change in the initial peak drag force. However, the ground effect does impact the parasitic thrust, which increases for all $\Xi$ values when the wing is close to the ground.  

To better understand our findings, we analyze the vorticity field for $\Xi$ = 0.2 \& 0.6 at two extreme ground heights, i.e., {$h^*$ = 1.5 \& 0.04}. Figure \ref{fig: Synchronous vorticity field} presents the normalized vorticity fields at time instances, $t^*$ = 0.37, 0.62, \& 0.85, which highlight the key changes in the flow field due to variations in $\Xi$ and ground height. To normalize the vorticity field, we use the wing's chord and the steady-state translational velocity, $\omega^*$ $=\frac{\omega * c}{U_\infty} $.

Our PIV results demonstrate that a rapid pitch-up motion during deceleration causes the shear layer to separate, leading to the formation of counter-rotating leading edge vortex (LEV) and trailing edge vortex (TEV) structures (Figure \ref{fig: Synchronous vorticity field}). Although both values of $\Xi$ result in similar vortex formation, smaller $\Xi$ produces larger LEV that diffuses faster and is farther away from the plate surface, whereas higher $\Xi$ leads to more coherent and stronger vortex structures closer to the wing. When the plate is pitching slowly while decelerating slowly, this motion generates a smaller pressure gradient on the wing surface, which creates weaker vortices that are more spread out. However, rapid pitch-up motion during rapid deceleration induces a large pressure gradient due to the rapid change in the flow direction and velocity, leading to the formation of stronger and more coherent vortex structures closer to the wing. The stronger and more coherent vortex closer to the wing surface induces more impulse on the wing than the vortices that are weaker and more spread out, explaining the larger value of lift and drag force observed for $\Xi$ = 0.6 compared to $\Xi$ = 0.2.

For both $\Xi$, at $t^*$ = 0.37 and 0.62, the TEV is larger and stronger at $h^*$ = 0.04 than at $h^*$ = 1.5, which is especially evident for $\Xi$ = 0.6. The proximity of the plate to the ground constrains the flow around the plate, leading to an increase of pressure below the wing \citep{ahmed_2005}, which causes the flow to curl more strongly around the edges of the plate \citep{lee_2018}, resulting in the larger and stronger TEV as seen in the near ground case. A stronger TEV, in turn, induces stronger velocities on the LEV, bringing them closer together. \citet{wu_1998} found that this close vortex pair induces a stronger downwash. For $\Xi$ = 0.6 at $t^*$ = 0.62, a stronger dipole jet oriented downward is generated in the near ground case, producing an upward force, which can explain the higher value of lift force observed at $h^*$ = 0.04 compared to $h^*$ = 1.5. 

The shed vortices near the ground experience additional shear stress from the ground due to the no-slip condition at the boundary. This shear stress prevents the shed vortices from spreading out, resulting in more compact shed vortices (see figure \ref{fig: Synchronous vorticity field} (d)). The compact shed vortices can reduce the adverse pressure gradient on the wing surface and delay the formation of secondary vortices, enhancing the stability of the LEV and increasing lift \textcolor{red}{\citep{mclean_2012}}. 

However, at the end phase of the maneuver, when the wing is close to the ground, the dipole jet gets impinged to the ground. At $t^*$ = 0.85, figure \ref{fig: Synchronous vorticity field} shows that this impingement advects the shed LEV and TEV further apart. For $\Xi$ = 0.6, the x-distance between the LEV and TEV is 0.92c for the near ground case compared to 0.76c for the far ground case, resulting in a lower impulse on the wing. This results in lower impulse on the wing. This may explain the increased drop in the lift and drag force at the end of the maneuver on the perching wing close to the ground.

\begin{figure}
\centerline{\includegraphics[width=1.0\textwidth]{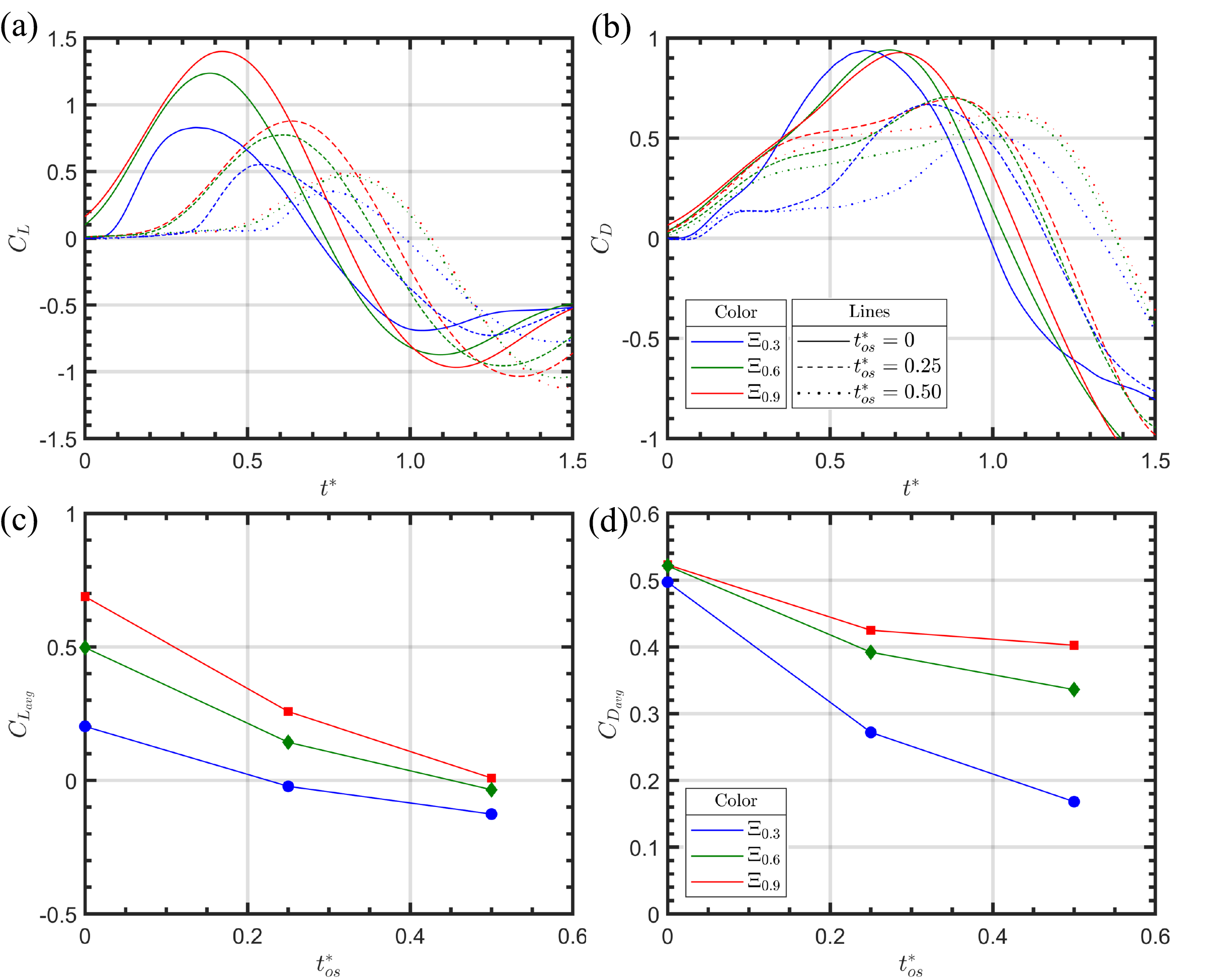}}
  \caption{Comparisons of the instantaneous (a) lift and (b) drag coefficient on the rectangular wing for $\Xi$ = 0.3, 0.6, and 0.9. Each $\Xi$ is presented at three-time offsets between the decelerating and pitch-up motion.}
\label{fig: Asynchronous motion lift and drag coefficient}
\end{figure}

\subsection{Asynchronous motion: Unsteady forces and flow-field}
\label{sec: Asynchronous motion results}
To investigate the influence of varying frontal area expansion on decelerating flight during perching, we employed asynchronous motion with three starting time offsets, $t^*_{os}$ = 0, 0.25, and 0.5 between the decelerating and pitch-up motions. Figure \ref{fig: Asynchronous motion lift and drag coefficient} (a) and (b) presents the evolution of lift and drag coefficient on the rectangular plate for three values of $\Xi$ and three values of $t^*_{os}$. Note that in the asynchronous case, the pitch rate is the same as in the synchronous case, but we reduced the deceleration value, resulting in an increase in $\Xi$ by a factor of \textcolor{red}{$\frac{1}{a}$}.

From figure \ref{fig: Asynchronous motion lift and drag coefficient} (a), it is observed that increasing $\Xi$ enhances the instantaneous lift coefficient, similar to the behavior observed in the synchronous case. For $t^*_{os}$ = 0, although the pitch rate is the same for both synchronous and asynchronous motions, the asynchronous motion produces a peak lift coefficient approximately $40\%$ higher due to the higher translational velocity experienced by the pitching plate. This higher velocity enhances both the circulatory and non-circulatory forces. However, for higher time offsets $t^*_{os}$ = 0.25 and 0.5, the lift coefficient starts to rise later and generates a lower peak lift coefficient compared to $t^*_{os}$ = 0. This delay in the rise of the lift coefficient is consistent with the delayed start of the pitch-up motion and the reduction in the lift coefficient can be correlated with the lower translational speed caused by the deceleration of the wing. These trends are consistent for all $\Xi$ considered in this study, with smaller $\Xi$ resulting in a lower peak lift coefficient.

Varying the starting time offsets between the deceleration and pitch-up motion leads to a distinct evolution of instantaneous lift coefficients. For $t^*_{os}$ = 0, the perching plate generates a high initial lift force but also experiences a rapid drop-off in the lift, which may reduce the control authority of landing birds. However, delaying the rapid pitch-up motion until late in the deceleration can delay the drop-off of lift force, allowing the wing to generate lift at the end of the motion and enhance control authority during this highly unsteady maneuver.    

The evolution of the drag coefficient for asynchronous motion is presented in figure \ref{fig: Asynchronous motion lift and drag coefficient} (b). At $t^*_{os}$ = 0, each value of $\Xi$ exhibits an increase in the peak drag coefficient compared to the synchronous case. As the starting time offset between the two motions increases, there is a reduction in the peak drag coefficient. Although the peak drag force is reduced at $t^*_{os}$ = 0.5, the perching plate generates a positive drag force for the majority of the flight, as opposed to negative drag or parasitic thrust in the latter stage of the maneuver for $t^*_{os}$ = 0. Furthermore, as we increase $t^*_{os}$, we observe a delay in the onset of parasitic thrust generation and a decrease in its magnitude. These results suggest that perching birds may have better control over the aerodynamic forces during landing at higher $t^*_{os}$ due to continuous drag generation and reduced parasitic thrust. 

\begin{figure}
\centerline{\includegraphics[width=1.0\textwidth]{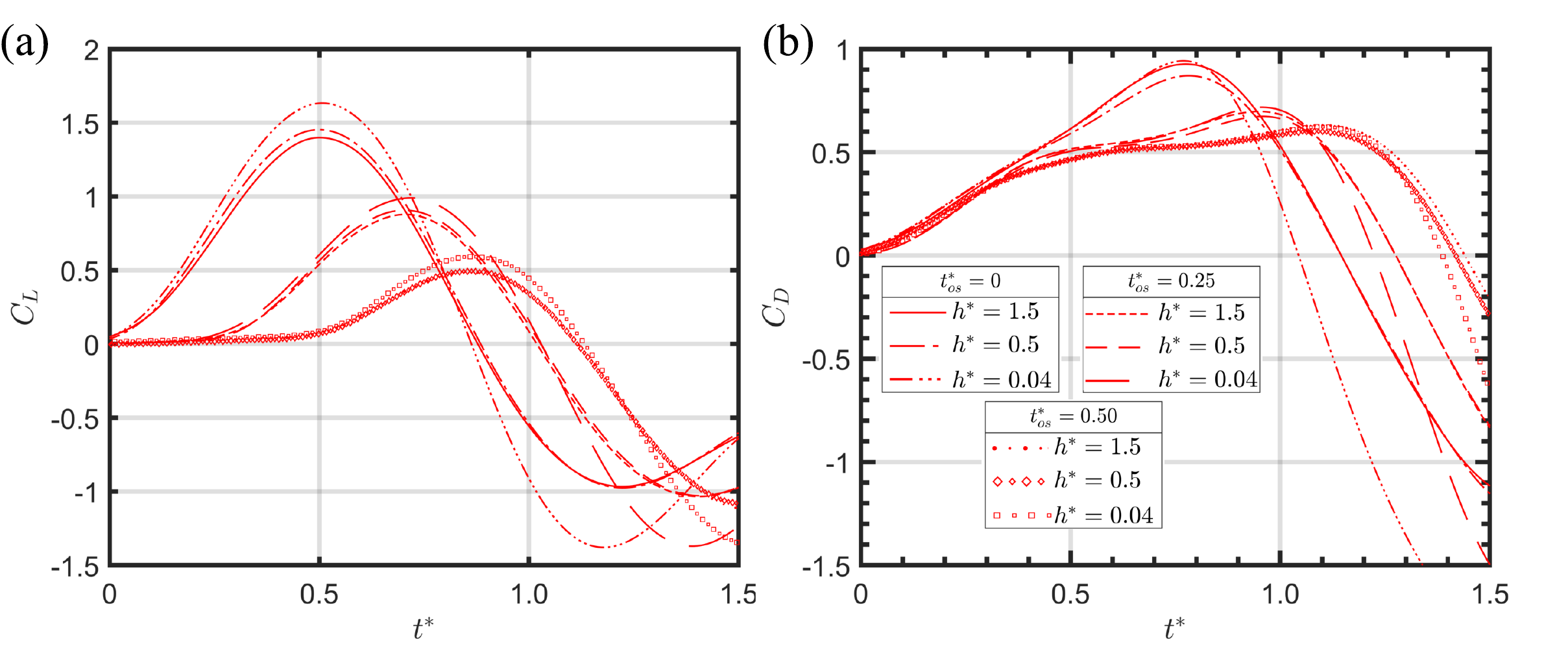}}
  \caption{Comparisons of the lift and drag coefficient on the rectangular wing for $\Xi$ = 0.9 at three non-dimensional ground height, $h^*$.}
\label{fig: Asynchronous motion lift and drag coefficient ground effect}
\end{figure}

We further analyzed the influence of asynchronous motion by plotting the time-averaged lift and drag coefficients in figure \ref{fig: Asynchronous motion lift and drag coefficient} (c \& d). Increasing the starting time-offset, $t^*_{os}$, results in a decrease in the time-averaged lift coefficient, $C_{L_{avg}}$ as shown in figure \ref{fig: Asynchronous motion lift and drag coefficient} (c). For $\Xi$ = 0.9, the $C_{L_{avg}}$ decreases from 0.7 to near zero as $t^*_{os}$ increases from 0 to 0.5. While the drag plot in figure \ref{fig: Asynchronous motion lift and drag coefficient} (d) also shows a reduction in the time-averaged drag coefficient, $C_{D_{avg}}$, with increasing $t^*_{os}$, the decrease in drag is small compared to that of the $C_{L_{avg}}$. For $\Xi$ = 0.9, the $C_{D_{avg}}$ decreases from 0.52 to 0.4 as $t^*_{os}$ increases from 0 to 0.5. The resulting near-zero lift force and positive drag force can help birds perch on the original landing or perching location without gaining altitude, providing a beneficial perching strategy.

Figure \ref{fig: Asynchronous motion lift and drag coefficient ground effect} shows the behavior of unsteady forces for $\Xi$ = 0.9 at three non-dimensional ground heights: $h^*$ = 1.5, 0.5, and 0.04. As the perching plate approaches the ground, the instantaneous lift coefficient increases similarly to synchronous cases, and this behavior is consistent across all starting time offsets. The result indicates that lower ground height provides more lift enhancement, with the greatest benefits observed at $t^*_{os}$ = 0. In contrast, ground proximity has less impact on drag force at the early stage of the maneuver (see figure \ref{fig: Asynchronous motion lift and drag coefficient ground effect}), but once the perching plate attains its peak at $t^*_{os}$ = 0, the drag force drops rapidly for near-ground case, creating higher negative drag force or parasitic thrust force at the end of the maneuver. While a similar drop in drag force is observed at $t^*_{os}$ = 0.5, this decrease in drag force is relatively small compared to the drop at $t^*_{os}$ = 0. This indicated that introducing a time offset could be the optimum way to execute a perching maneuver during landing, as it helps reduce the risk of losing control authority over aerodynamic forces. 

\begin{figure}
\centerline{\includegraphics[width=1.0\textwidth]{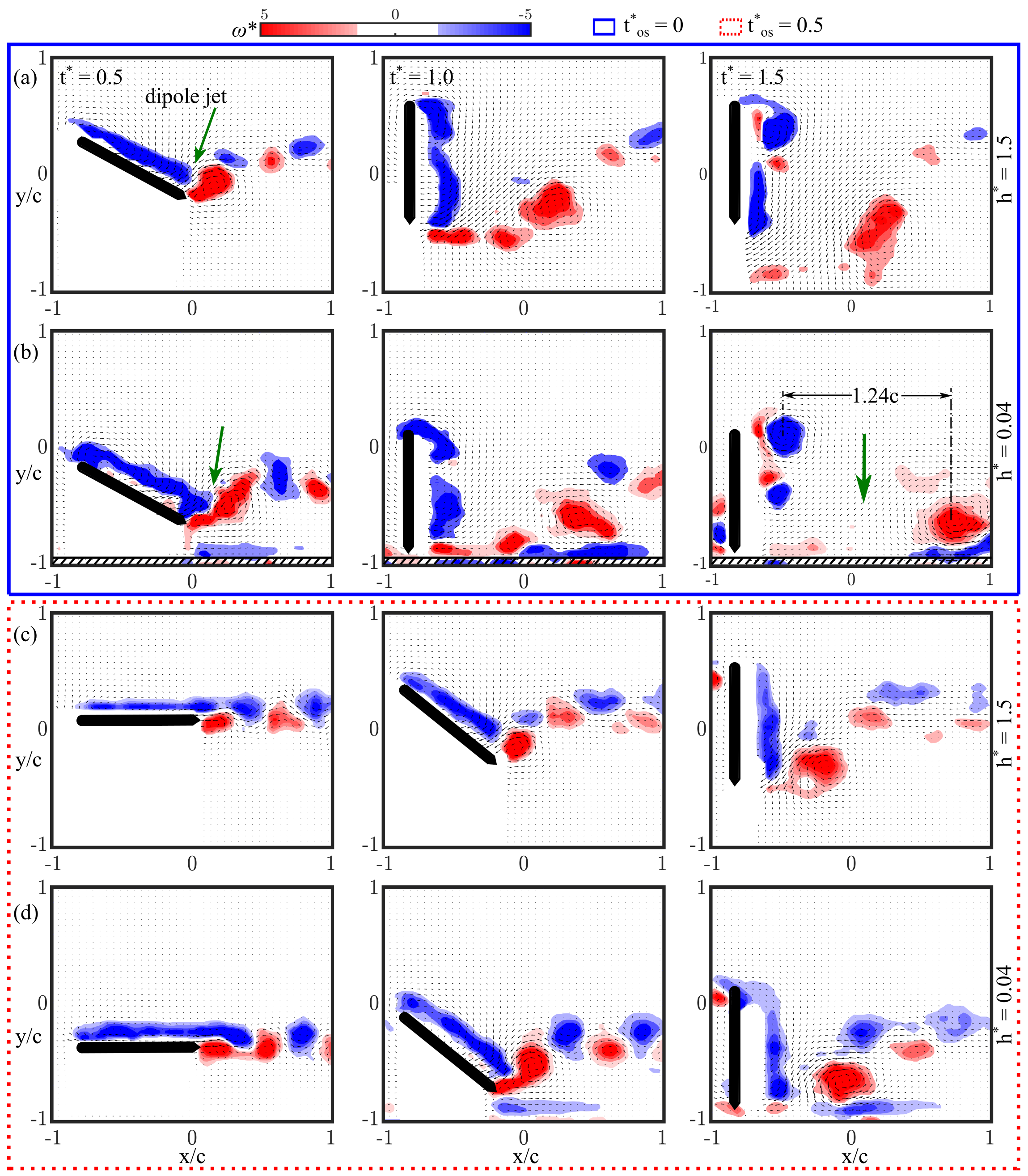}}
  \caption{Contours of the normalized vorticity field, $\omega^*$, for asynchronous motion at the $50\%$ of the wing span for $\Xi$ = 0.9 at three time-steps, $t^*$ = 0.5, 1.0, \& 1.5: $t^*_{os}$ = 0 at (a) $h^*$ = 1.5; (b) $h^*$ = 0.04. $t^*_{os}$ = 0.5 at (c) $h^*$ = 1.5; (d) $h^*$ = 0.04.}
\label{fig: Asynchronous vorticity field}
\end{figure}

We investigated the normalized vorticity field to further analyze the results observed in the asynchronous motion. Specifically, we focused on the highest shape change number, $\Xi$ = 0.9, and examined two starting time-offset cases, $t^*_{os}$ = 0 \& 0.5, corresponding to executing rapid pitching at different deceleration stages. This allowed us to highlight the key change in the evolution of flow physics in asynchronous motion. Figure \ref{fig: Asynchronous vorticity field} displays the normalized vorticity field at three-time instants, $t^*$ = 0.5, 1.0, and 1.5, and at two extreme ground height cases, $h^*$ = 1.5 and 0.04.

Examining figure \ref{fig: Asynchronous vorticity field} for $t^*_{os}$ = 0, we observed that at $t^*$ = 0.5, the pitching plate generates a coherent leading-edge vortex (LEV) and trailing-edge vortex (TEV). These stronger and more coherent vortex structures produce higher lift and drag forces, which is consistent with the evolution of higher unsteady forces observed in the instantaneous force plot for $t^*_{os}$ = 0. However, at the same time instant, the vorticity field for $t^*_{os}$ = 0.5 reveals the development of the wake downstream of the plate. This wake opposes the motion of the wing and contributes to the generation of drag force, explaining the early stage generation of drag force in $t^*_{os}$ = 0.5 cases. Furthermore, as seen in the synchronous case, the vortices are stronger at $h^*$ = 0.04 than at $h^*$ = 1.5, explaining the higher lift force observed at $h^*$ = 0.04 compared to $h^*$ = 1.5. 

Meanwhile, at $t^*$ = 1.0, the vortices are fully developed and shed from the wing surface for $t^*_{os}$ = 0 cases, but for $t^*_{os}$ = 0.5 cases, the vortices are still growing. This difference in the vortex development explains why the unsteady forces drop for $t^*_{os}$ = 0 cases but continue to increase for $t^*_{os}$ = 0.5 cases at $t^*$ = 1.0. In the former case, stronger and more coherent vortex structures developed earlier in the maneuver induce a stronger dipole jet, which impinges on the grounds and separates the shed vortices further apart, as seen in figure \ref{fig: Asynchronous vorticity field} (b) at $t^*$ = 1.0 and $t^*$ = 1.5. At $t^*$ = 1.5, for $t^*_{os}$ = 0 case close to the ground, the x-distance between the LEV and TEV is \textcolor{red}{1.24c}, contributing to the more pronounced drop-off of the unsteady force. In contrast, for $t^*_{os}$ = 0.5 cases, the weaker vortices result in a slower jet and closer separation between the shed vortices (\textcolor{red}{0.6c}) and the wing surface, resulting in a less pronounced drop-off of the unsteady force.

\subsection{Discussion on Dipole Jet}\label{sec: Discussion on Dipole Jet}

\begin{figure}
\centerline{\includegraphics[width=1.0\textwidth]{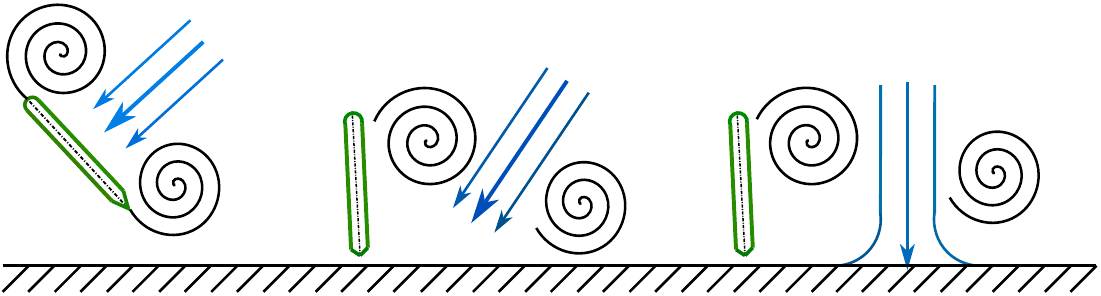}}
  \caption{Schematics showing the evolution of the dipole jet and its interaction with the ground.}
\label{fig: Dipole Jet}
\end{figure}

This research aims to investigate the mechanics of rapidly pitching plates during deceleration near the ground to understand how the perching and hunting birds achieve their impressive maneuverability. By executing rapid pitching at different stages of deceleration, such as when the forward translational velocity is still high versus when it is low, we observe distinct changes in the evolution of vortex structures and forces acting on the plate. 

Executing a rapid pitch-up motion during deceleration causes the shear layer to separate, forming a counter-rotating leading edge vortex (LEV) and trailing edge vortex (TEV) vortex structures, as shown by our PIV results. The pitch-up motion combines the counter-rotating vortices to form a comoving vortex dipole. Once a vortex dipole is created, they will interact with each other, creating a region of high vorticity gradient. This region acts as a source of fluid, moving fluid outward, supplying fluid to the vortices needed to maintain their coherent structures, and forming the jet flow of the dipole. At the same time, the region between the two vortices experiences a low-pressure zone due to the centrifugal forces generated by the counter-rotating vortices. This low-pressure region acts as a sink flow, drawing fluid between the two vortices. This combination of source flow and sink flow forms a dipole jet that is characteristic of counter-rotating vortices. The schematic diagram in figure. \ref{fig: Dipole Jet} illustrates the pitch-up motion of the wing and the resulting formation of the dipole jet. This dipole jet moves a considerable amount of momentum carrying fluid with it, which can be used to generate unsteady forces on the wing. 

\begin{figure}
\centerline{\includegraphics[width=1.0\textwidth]{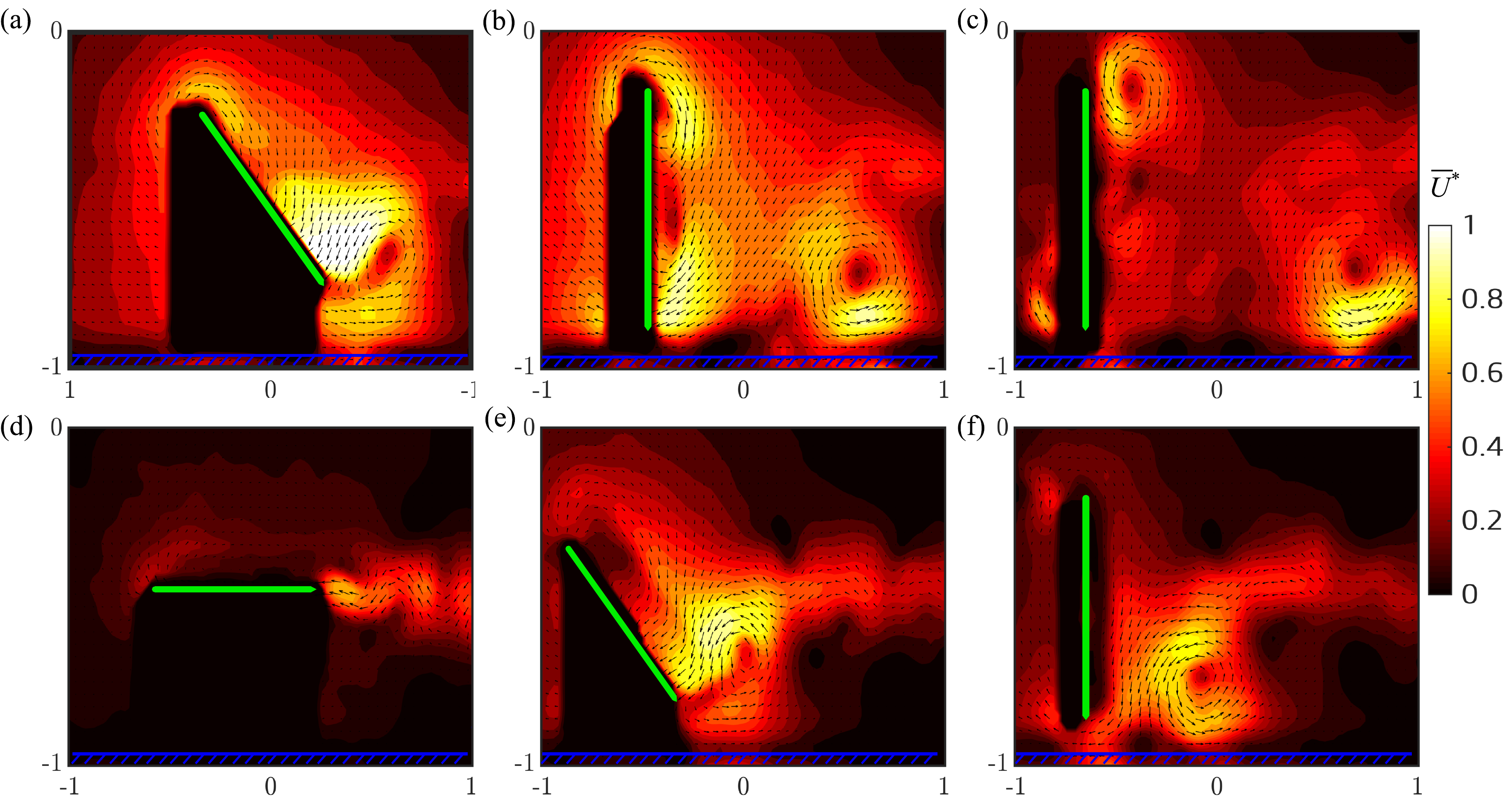}}
  \caption{Velocity field for close to ground case, $h^*$ = 0.04. (a, b, c) $t^*_{os}$ = 0; (d, e, f) $t^*_{os}$ = 0.5. (a, d) $t^*$ = 0.5; (b, e) $t^*$ = 1; (c, f) $t^*$ = 1.5.}
\label{fig: Dipole Jet asymetric case}
\end{figure}

Figure \ref{fig: Dipole Jet asymetric case} displays the dipole jet formation during the asynchronous motion by analyzing the velocity field at two different starting time-offsets cases, $t^*_{os}$ = 0 \& 0.5. For $t^*_{os}$ = 0 case, pitch-up motion is executed at a high forward translational velocity, resulting in larger, stronger counter-rotating vortices and a more significant induced dipole jet (as seen in figure \ref{fig: Dipole Jet asymetric case} (a)). This dipole jet is initially directed downward and forward, producing lift and drag forces. However, when the wing pitches up to its final effective angle of attack, i.e., at $t^*$ = 1, the dipole jet gets deflected on the ground surface. This deflected jet moves the vortex pair apart, reducing the impulse they generate on the wing and causing a rapid drop in the lift and drag force at the end phase of the maneuver. At $t^*$ = 1.5, some deflected jets are reversed and oriented backward, producing a parasitic thrust force. \citet{weymouth_2013} showed that to achieve the ultra-fast escape, the deforming body stores added mass energy in the fluid during the early phase of the maneuver by deforming. This energy is then recovered in the later phase of the maneuver to accelerate the deformed body. In the present study, the pitching plates recover the deflected jet, which is oriented backward in the later phase of the motion, as parasitic thrust. Hunting birds like eagles can use this parasitic thrust to accelerate after catching their prey.

On the other hand, when executing a pitch-up motion at a low forward translational velocity ($t^*_{os}$ = 0.5), smaller and weaker vortex structures are formed, which induces a slower dipole jet that moves a lower amount of momentum-carrying fluid with it(see figure \ref{fig: Dipole Jet asymetric case} (e)). This motion generates vortices, and the dipole jet at the end phase of the maneuver, providing lift and drag force suitable for a smooth touch-down. As this motion induces a slower dipole jet, it does not displace the vortex pair further apart upon impact with the ground, reducing its influence on the rapid drop-off of the lift and drag force compared to $t^*_{os}$ = 0 scenario. 

We correlated the induced dipole jet with increased added mass during the frontal area expansion. The one-dimensional added mass force for an expanding body can be rewritten as:

\begin{equation}
  F_{AM}=-\frac{\partial}{\partial t} (m_{a}U) = -\dot m_{a}U - m_{a}\dot U
  \label{added mass parasitic thrust}
\end{equation}

From this equation, for expanding body, the addition of added mass from frontal area change creates drag from $-\dot m_{a}U$. However, this frontal area expansion also increases total added mass, making the body difficult to stop at the latter stage of the maneuver due to the generation of the net thrust through $-m_{a}\dot U$.

In asynchronous motion, $t^*_{os}$ = 0 case induces a larger, stronger dipole jet, increasing the added mass early in the maneuver. This increased added mass generates higher parasitic thrust through $-m_{a}\dot U$ later in the maneuver, which the bird can use to accelerate after catching the prey. Conversely, when the frontal area expansion occurs at the low velocity (i.e., $t^*_{os}$ = 0.5), a slower dipole jet is induced, leading to a smaller increment in the added mass and a lower value of parasitic thrust, resulting in smooth touch down during landing or perching.

In summary, this study demonstrates the importance of dipole jets in achieving specific flying objectives. We observed that by directing the dipole jet in different directions, birds could generate lift, drag, or thrust force to suit their needs during landing, perching, or hunting. These findings provide new insight into the intricate relationship between the dipole jet and the added mass forces, offering a new perspective on the performance of the bird's flight and the design for safer flying vehicles.

\begin{figure}
\centerline{\includegraphics[width=1.0\textwidth]{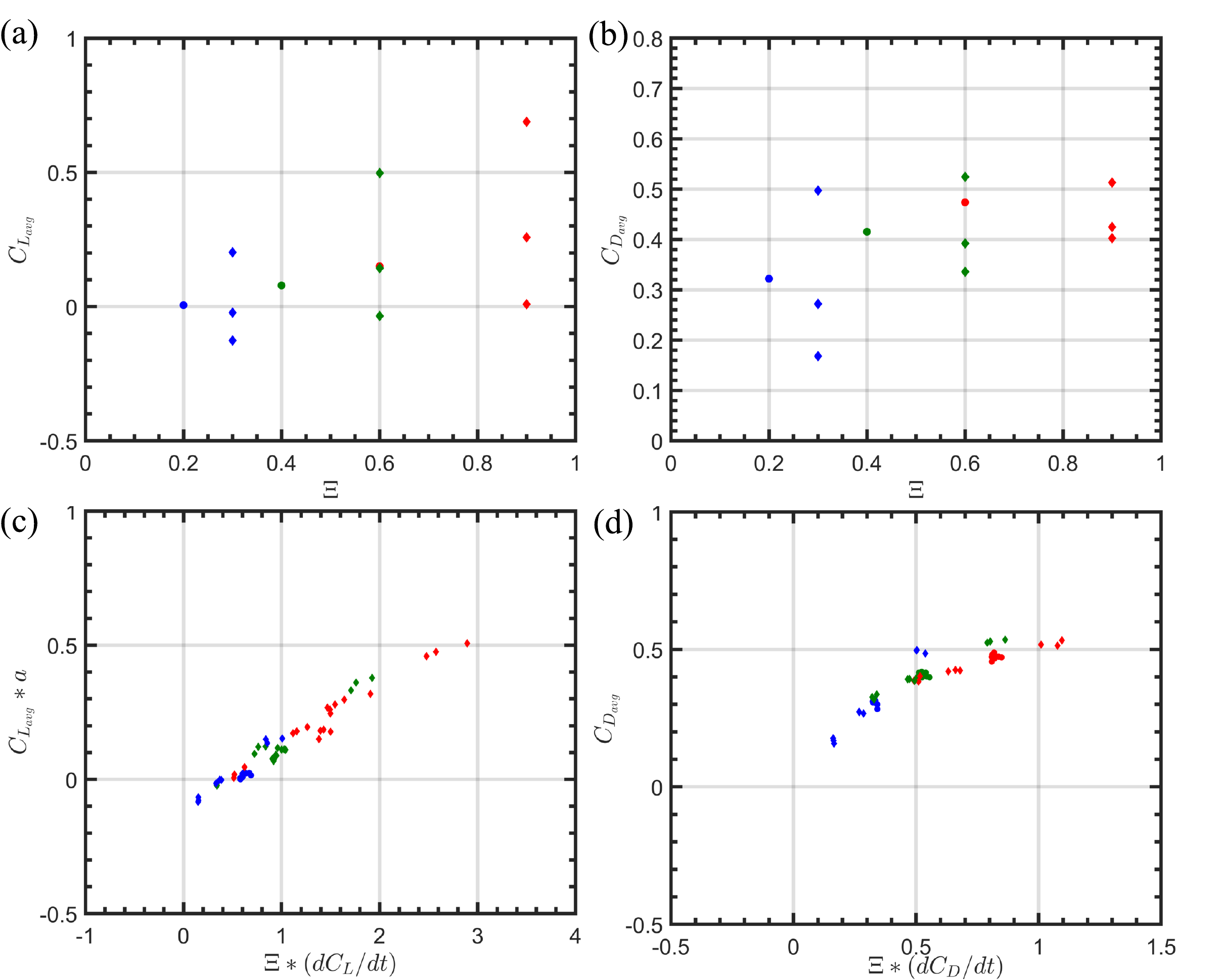}}
  \caption{(a) Lift and (b) drag coefficient on a finite rectangular wing at the non-dimensional ground height of $h^*$ = 1.5 as a function of $\Xi$. Scaling of the (c) lift and (d) drag coefficients for all the $\Xi$ and ground heights considered in the experiment. $C_{L_{avg}}$ is multiplied by a factor of a, which is the ratio of the pitch time period to the deceleration time period.}
\label{fig: Scaling lift and drag coefficient}
\end{figure}

\subsection{Scaling Laws}\label{sec: Scaling laws results}
The total lift and drag forces generated during the rapid pitching motion are crucial for achieving specific flying objectives. We analyze time-averaged lift and drag coefficients across various rapid pitching scenarios to evaluate flight success. We developed a new scaling law to describe the performance of the rapidly pitching plates in decelerating motion near the ground. 

In synchronous motion at a non-dimensional ground height of $h^* = 1.5$, we observed a linear increase in the time-averaged forces with increasing $\Xi$ (for reference, see figure \ref{fig: Scaling lift and drag coefficient} (a) \& (b) represented by the symbol, \scalebox{0.6}{$\CIRCLE$}). This finding is consistent with a previous study by \citet{polet_2015}. However, in asynchronous motion, we found that executing pitch-up motion at different stages of deceleration generated a more complex relation between time-averaged forces and $\Xi$ (represented by the symbol, \scalebox{0.6}{$\blacklozenge$} in figure \ref{fig: Scaling lift and drag coefficient} (a) \& (b)). Specifically, we found that increasing the starting time offsets, $t^*_{os}$, from 0 to 0.5, at the same $\Xi$, led to a decrease in time-averaged lift and drag forces. 

To account for this complex relationship, we introduced new scaling relations by considering the rate of change of maximum lift and drag coefficient, $\frac{dC_{L_{max}}}{dt}$ \& $\frac{dC_{D_{max}}}{dt}$, which were found to be inversely proportional to the starting time offsets, $t^*_{os}$. We multiplied this rate of change with the $\Xi$ and the averaged lift coefficient ($C_{L_{avg}}$) by a factor of a, which is the ratio of pitch to deceleration time period. We used this scaling law to scale the time-averaged lift and drag coefficient for all tested scenarios, and our experimental data demonstrate good agreement with this new scaling law (see figure \ref{fig: Scaling lift and drag coefficient} (c) \& (d)).

The main improvement of this scaling law is the connection between the execution of the pitch-up motion during decelerating motion and the generation of total lift and drag forces. Our findings indicate that the timing of pitch-up motion influences the evolution of the flow field around the plates, leading to changes in the development of lift and drag forces. This scaling law can be applied to all pitching plates in decelerating motion, regardless of the ground height, as shown in figure \ref{fig: Scaling lift and drag coefficient} (c) \& (d). This modification simplifies the prediction of unsteady forces on rapidly pitching plates in decelerating motion, enabling accurate prediction across a broad range of parameter space. 

\subsection{Modeling the variation of unsteady forces during a perching maneuver}\label{sec: Analytical model and results}
The perching wing experiences two main types of forces: added mass force and circulatory force. 

Added mass force, also known as non-circulatory force, arises due to the acceleration of the fluid during the unsteady motion of the wing. For a rapidly pitching plate in decelerating motion, the added mass force coefficient can be determined using the following equation:

\begin{equation}
    C_{F_{nc}} = \frac{\pi c}{2U^2_{\infty}}[\dot{\alpha}cos(\alpha)U + sin(\alpha)\dot{U} + c\ddot{\alpha}(1/2 - x^*_p)]
    \label{eq: added mass force coefficient}
\end{equation}

The first term in this equation denotes the rate of change of added mass, while the second and third term corresponds to the added mass due to the accelerating or deceleration of the wing and rotational acceleration, $\ddot{\alpha}$, respectively. Here, the pivot location on the wing is at the mid-chord, resulting in the relative distance of $(1/2 - x^*_p)$ equal to zero. When a body is close to the ground, the acceleration of the fluid between the wing and the ground increases leading to a substantial increase in the added mass force \citep{brennen_1982}. To account for this effect, we modified the added mass equation as:

\begin{equation}
    C_{F_{nc}} = \frac{\pi c}{2U^2_{\infty}}[\dot{\alpha}cos(\alpha)U + sin(\alpha)\dot{U} + c\ddot{\alpha}(1/2 - x^*_p)][1+k(1/2h^*)^2]
    \label{eq: modified added mass force coefficient}
\end{equation}

\noindent where k is the constant determined experimentally, with k = 0.002.

In this study, we developed an analytical model to calculate the unsteady forces acting on a finite wing as it rapidly pitches and decelerates near the ground. We incorporated the finite wing and unsteady effect by using a combination of Wagner's theory and the unsteady lifting line model (LLT) following \citet{boutet_2018}. Here, we modified the LLT to include the ground effect behavior using image vortices.   

To represent the flat plate and the trailing vortex sheet, we used the lifting line approach. From Prandtl's lifting line theory, the downwash velocity, $w_y$, induced at spanwise location $y$ is given by:

\begin{equation}
     w_{y} = \frac{-1}{4\pi}\int_{-S/2}^{S/2}\frac{\frac{d\Gamma}{dy_0}}{y - y_0}{dy_0},
    \label{eq: downwash due to trailing vortex}
\end{equation}

Here, $\Gamma$ is the strength of the vortex. For the unsteady case involving a finite wing, $\Gamma$ is a function of time ($t$) and span location ($y$) distribution and can be represented by a Fourier series with a time varying Fourier coefficients, $a_n$, as: 

\begin{equation}
    \Gamma(t,y) = \frac{1}{2}a_0 c_0 U\sum_{n=1}^{N}a_n(t)\sin(n \theta),
    \label{eq: time-dependent Fourier series representation of the circulation}
\end{equation}

\noindent where $a_0$ is the lift curve slope, with $a_0 = 2\pi$ for an ideal airfoil, $c_0$ is the chord length of the wing, and $N$ is the number of spanwise stripes. Here the coordinate transformation of $y = \frac{s}{2}cos(\theta)$ is applied to map the angle, $\theta$, to the semi-span, $\frac{s}{2}$, position of the wing. Substituting Eq.(\ref{eq: time-dependent Fourier series representation of the circulation}) into Eq.(\ref{eq: downwash due to trailing vortex}) and applying the Glauert integral \citep{glauert_1983} leads to a simplified form of downwash velocity as:

\begin{equation}
    w_y (t) = -\frac{a_0 c_0 U}{4S}\sum_{n=1}^{N} n a_n(t)\frac{\sin(n \theta)}{sin(\theta)},
    \label{eq: quasi-steady version of the lifting line theory}
\end{equation}

To include the ground effect, we incorporate an image vortex system that satisfies the zero normal flow boundary condition on the ground. This image vortex system induces an upwash on the finite wing, effectively modifying the tip vortex-induced downwash velocity. The upwash velocity, $w_{I_y}$, induced by the image lifting line can be represented following \citep{ariyur_2005}:

\begin{equation}
    w_{I_y}(t) = \frac{cos(2\beta)}{4\pi}\int_{-S/2}^{S/2}\frac{(y - y_0)\frac{d\Gamma (t)}{d y_0}\quad{dy_0}}{[(y - y_0)^2 + 4h^2cos^2(\theta)]},
    \label{eq: upwash due to image lifting line}
\end{equation}

This upwash velocity is the main contribution of the ground effect, which alters the lift forces on the plate. After coordinate transformation and substitution of the derivative of $\Gamma$ into Eq. (\ref{eq: upwash due to image lifting line}), the upwash velocity can be expressed as:

\begin{equation}
    w_{I_y} (t) = \frac{cos(2\beta)}{\pi}\\\int_{0}^{\pi}\frac{(cos(\theta_0) - cos(\theta))\sum_{n=1}^{m} n a_n cos(n \theta_0)} {[(cos(\theta_0) - cos(\theta))^2 + 16(\frac{h}{b})^2cos^2(\beta)]}{d\theta_0},
    \label{eq: transformed upwash velocity}
\end{equation}

The unsteady sectional lift coefficient, $c_l^c (t)$, can be obtained in terms of $a_n(t)$ using the unsteady Kutta-Joukowski theorem \citep{katz_2001}:
\begin{equation}
    c_l^c (t) = \frac{2\Gamma}{U c(y)} + \frac{2\dot \Gamma}{U^2},
    \label{eq: circulatory sectional lift coefficient}
\end{equation}

By substituting the Fourier series representation of the vortex strength (Eq.\ref{eq: time-dependent Fourier series representation of the circulation}) for $\Gamma$, $c_l^c (t)$ can be expressed as:
\begin{equation}
 c_l^c (t) = a_0 \sum_{n=1}^{N}(\frac{c_0}{c} a_n + \frac{c_0}{U}{\dot a_n})\sin(n\theta),
\label{eq: Fourier circulatory sectional lift coefficient}
\end{equation}

The unsteady motion of the flat plate causes a step change in the downwash velocity, $\Delta w(y)$, along the span. The variation of the circulatory lift coefficient due to this step change is given in terms of indicial function:
\begin{equation}
    c_l^c(t) = a_0(y)\Phi(t) \frac{\Delta w(y)}{U},
    \label{eq: Wagner sectional lift coefficient}
\end{equation}

Here, $\Phi(t)$ represents the Wagner function, which can be expressed as:  
\begin{equation}
    \Phi(t) = 1 - \Psi_1 e^-\frac{\epsilon_1 U}{b}t - \Psi_2 e^-\frac{\epsilon_2            U}{b}t,
    \label{eq: Jones approximation of Wagner Function}
\end{equation}

\noindent where the constants $\Psi_1$ = 0.165, $\Psi_2$ = 0.335, $\epsilon_1$ = 0.0455 and $\epsilon_2$ = 0.3 are derived from Jone's approximation of the Wagner function \citep{jones_1938}.

To capture the continuous lift response of an airfoil that undergoes arbitrary motion, we use Duhamel's integral. This method involves superimposing the step response of the Wagner function $\Phi(t)$ with the differential variation of the downwash velocity, $w(t,y)$. However, in our case, the flat plate experiences gradual deceleration, and the free-stream velocity decreases with time. To account for this variation, we modify Duhamel's integral formulation by considering $U = U(t)$, as suggested by \citet{van_1992} and \citet{hansen_2004}. Based on this, we rewrite Duhamel's integral formulation with the time-varying free stream velocity as:

\begin{equation}
    c_l^c(s) = \frac{a_0}{U}\bigg(w(0)\Phi(s)+ \int_{0}^{s}\frac{(\partial w(\tau)}{\partial \tau}\Phi(s-\tau)d\tau\bigg),
    \label{eq: Wagners circulatory lift coefficients}
\end{equation}
    
\noindent where $s$ is the non-dimensional time scale, which is calculated as $s = \frac{2}{c}\int_{0}^{t}Udt$.

Applying integration by parts to Duhamel's integral and substituting the first order differential equation derived by differentiating the two time lags terms of the $\Phi(t)$ with respect to $t$, yields the following expression of the sectional circulatory lift coefficient along the wing span:

\begin{equation}
    c_l^c(t,y) = \frac{a_0(y)}{U}(w(t,y)(1-\Psi_1-\Psi_2) + y_1(t)+y_2(t)
    \label{eq: effective circulatory lift coeff}
\end{equation}

\noindent where the state variables $y_i$ are defined as:
\begin{equation}
    y_i(t) = \Psi_i\epsilon_i\frac{2}{c}\int_{0}^{t}w(t',y)U(t')e^{-\epsilon_i\frac{2}{c}\int_{t'}^{t}U(\tau)d\tau}dt'
    \label{eq: state variables}
\end{equation}

In this study, the downwash on the wing is caused by both the motion of the wing and the 3D wake. The motion of the wing contributes to the downwash through pitch and angle of attack. The 3D downwash is calculated using modified lifting line theory, which considers both downwash from the trailing vortex sheet and upwash from the image vortex sheet. The total downwash $w(t,y)$ on the wing is expressed as: 
\begin{equation}
    \textcolor{red}{w(t, y) = U\alpha_y(t) + \dot\alpha_y(t)d +w_y(t) + w_{I_y}(t)}
    \label{eq: downwash on the wing}
\end{equation}
where d represents Theodersen's nondimensional distance.

To remove the integrals from the state variables, new state variables, $z_k$, are introduced:
\begin{equation}
        z_k(t, y) = \int_{0}^{t} e^{-\epsilon_i\frac{2}{c}\int_{t'}^{t}U(\tau)d\tau}v_k(t', y)dt',\quad i = 1,2
        \label{eq: Aerodynamic state variable}
\end{equation}

\noindent where k = 1,2,...6, $v_{1,2}$ = $\alpha$, $v_{3,4}$ = $\frac{w_y}{U}$ and $v_{5,6}$ = $\frac{w_{I_y}}{U}$. Here, i = 1 is used for k = 1, 3, and 5, and i = 2 for k = 2, 4, and 6. 

To express the first-order differential equation of Eq. \ref{eq: Aerodynamic state variable}, we used Leibniz's integral rule:
\begin{equation}
        \dot z_k(t, y) = v_k - \frac{\epsilon_iU}{b}z_k(t, y),\quad i = 1,2
 \end{equation}

 After combining Eqs.(\ref{eq: Fourier circulatory sectional lift coefficient}), (\ref{eq: effective circulatory lift coeff}) and (\ref{eq: downwash on the wing}), and performing the substitution of $z_k$ into $y_i$, we can now turn Eq. (\ref{eq: effective circulatory lift coeff}) at the $j_{th}$ stripe into a Wagner lifting line matrix equation as:

 \begin{equation}
    \begin{aligned}
        \boldsymbol{D_{yj}}\boldsymbol{\dot a_n} &= \boldsymbol{J_{j}} \dot {\textbf{p}} + \boldsymbol{K_{j}}\textbf{p} + \boldsymbol{L_{j}}\textbf{z} + (r(y)(\boldsymbol{W_{yj}} + \boldsymbol{W_{Iyj}})-\boldsymbol{A_{yj}})\boldsymbol{a_n} \\
        \dot {\textbf{z}} &= \boldsymbol{E_{j}}\textbf{z} + \boldsymbol{F_{j}}\textbf{p} + \frac{\textbf{G}}{U}(\boldsymbol{W_{yj}} + \boldsymbol{W_{I_yj}})\boldsymbol{a_n}
        \label{eq: Wagner Lifting Line theory}
    \end{aligned}
 \end{equation}

\noindent where,
\begin{flalign*}
    \textbf F_j &= [1\ 1\ 0\ 0\ 0\ 0]^T,&&\\
    \textbf D_{yj} &= a_0 \sum_{n=1}^{N}\frac{c_0}{U}\sin(n\theta_j),&&\\
    \textbf A_{yj} &= a_0 \sum_{n=1}^{N}\frac{c_0}{c}\sin(n\theta_j),&&\\ \textbf E_j &= \frac{-U}{b} diag (\epsilon_{i=1},\quad  \epsilon_{i=2},...),&&\\ 
    \textbf{z} &= [z_{k=1}\quad z_{k=2}\quad ... ]^T,&&\\
    \textbf{G} &= [0\ 0\ 1\ 1\ 0\ 0;\ 0\ 0\ 0\ 0\ 1\ 1]^T,&&\\
    \textbf{p} &= [\alpha(t,y)],&&\\ 
    \textbf J_j &= \frac{a_0(y)}{U}\Phi(0)d,&&\\
    \textbf K_j &= \frac{a_0(y)}{U}[U\Phi(0) + d\dot\Phi(0)],&&\\ 
    \textbf L_j &= \frac{a_0(y)U}{b}[\Psi_1\epsilon_1(1-\epsilon_1\frac{d}{b})\quad \Psi_2\epsilon_2(1-\epsilon_2\frac{d}{b})\quad \Psi_1\epsilon_1\quad \Psi_2\epsilon_2\quad \Psi_1\epsilon_1\quad \Psi_2\epsilon_2]^T, and &&\\
    r(y) &= \frac{a_0(y) \Phi(0)}{U} &&.
\end{flalign*}

In this study, the wing is divided into $N$ spanwise sections. When we apply Eq.(\ref{eq: Wagner Lifting Line theory}) to all the N sections, a system of 9N differential equations is formed. A MATLAB-built ODE function, ode15s, is used to solve the given systems of ODE. After computing the Fourier coefficients, $a_n$, Eq.(\ref{eq: circulatory sectional lift coefficient}) is used to calculate the lift coefficient along the wing span.

\begin{figure}
\centerline{\includegraphics[width=1.0\textwidth]{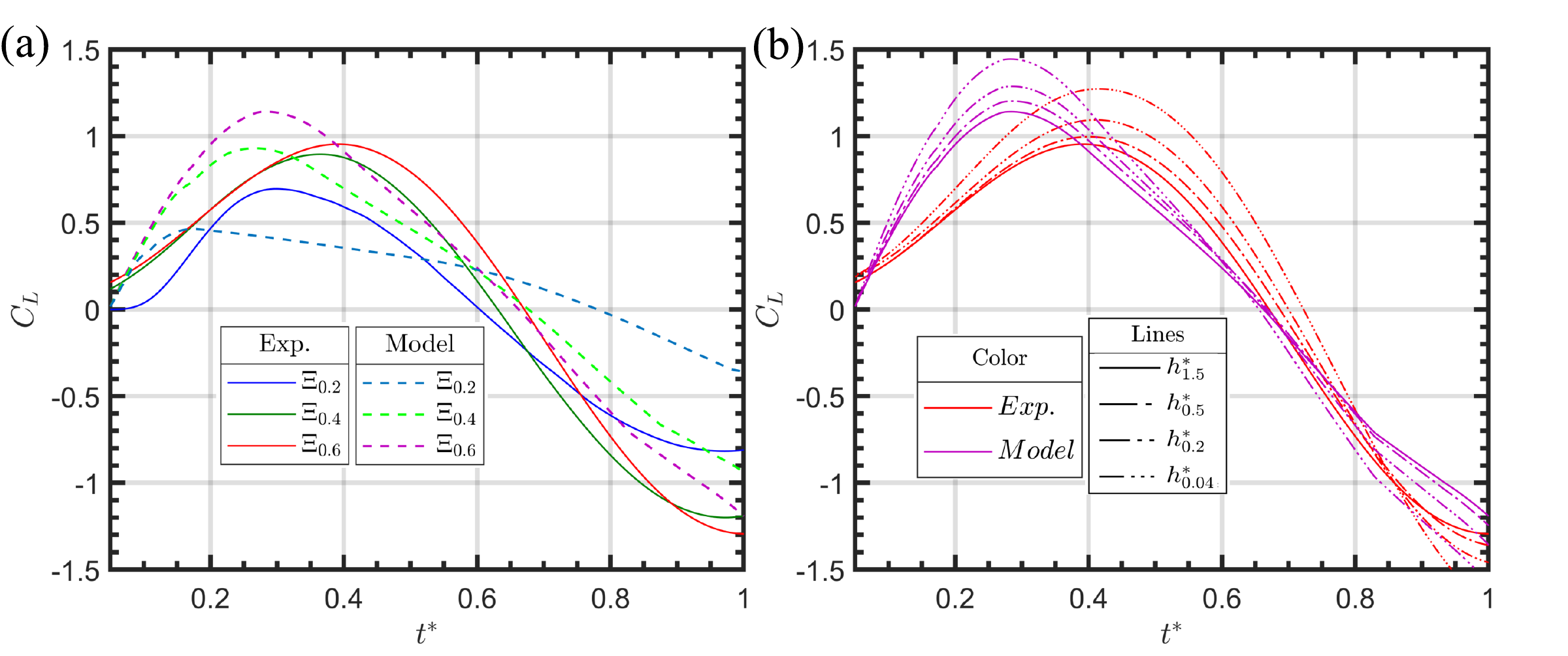}}
  \caption{Comparison of lift coefficients on the flat plate: (a) Experimental and predicted results for three values of $\Xi$ at $h^*$ = 1.5; (b) Experimental ground effect data and predicted results for $\Xi$ = 0.6.}
\label{fig: Comparison Experimental data with Predicted results}
\end{figure}

Figure \ref{fig: Comparison Experimental data with Predicted results} compares the time-varying lift coefficient, $C_L$, predicted by our analytical model with the experimental results. In figure \ref{fig: Comparison Experimental data with Predicted results} (a), a steep rise in $C_L$ is observed during the initial deceleration phase for all values of $\Xi$, consistent with the experimental findings. This suggests that both the non-circulatory and the circulatory forces play a dominant role in generating unsteady forces in pitching plates during deceleration. Subsequently, $C_L$ decreases, as observed experimentally, due to a reduction in non-circulatory and circulatory forces. The analytical model successfully captures the general trend of the instantaneous lift coefficient evolution observed in the experimental results (Figure \ref{fig: Comparison Experimental data with Predicted results} (a)). However, accurate prediction of lift evolution in pitching plates requires a precise estimation of vortex growth and detachment from the wing surface, which the present analytical model lacks. This discrepancy explains the disparity between the predicted and experimental $C_L$ values. 

Figure \ref{fig: Comparison Experimental data with Predicted results} (b) compares the ground effect results obtained from experiments with the predicted results from analytical model for $\Xi$ = 0.6. To incorporate the ground effect, we modified the added mass equation and integrated an image vortex system into the circulatory force analysis. With this modification, the present analytical model successfully predicts the observed increase in the peak $C_L$ as the wing approaches the ground and captures the corresponding drop in $C_L$ and increase in negative lift as the wing gets closer to the ground. However, there is a slight deviation between the predicted peak $C_L$ and the experimental values for all ground heights, primarily due to model's limited ability to accurately estimate vortex growth and detachment from the wing surface. Nevertheless, the present model effectively captures the general trend in the evolution of $C_L$ and the influence of ground effect on rapidly pitching plates during deceleration.

\section{Conclusion}
We used a rapid pitch-up motion of a finite flat plate in a decelerating flow as a simple model to investigate the effect of frontal area change on unsteady forces and flow physics in perching or hunting birds. Experiments were conducted to measure the instantaneous force and flow development at different pitch rates and decelerations. We tested the same kinematics at different ground heights to explore the unsteady ground effect experienced by the rapidly pitching birds. We developed a simple analytical model that incorporates the added mass forces and the circulatory force to predict the instantaneous lift force on the wing. The predicted results were compared with the experimental data to validate the accuracy of the model. 

To safely land or perch, birds need enough lift force to support their weight and drag force to decelerate to a complete stop. Our research shows that in synchronous motion increasing the shape change number $\Xi$ leads to higher lift and drag forces on the wing. This higher unsteady force is the outcome of the combined increase in non-circulator and circulatory force resulting from the higher pitch rate. While this higher drag force is crucial for the rapid dissipation of the kinetic energy during landing or perching, however, this higher lift force may prevent the bird from landing at the intended location or altitude. 

In asynchronous motion, varying the starting time offsets between the deceleration and pitch-up motion results in a distinct evolution of instantaneous lift and drag forces. For the starting time offset, $t^*_{os}$ = 0, the plate generates a high initial lift and drag force but experiences a rapid drop-off of the aerodynamic forces later in the motion. However, delaying the rapid pitch-up motion until late in the deceleration (i.e., $t^*_{os}$ = 0.5) allows the wing to generate lift and drag at the end of the motion. In this case, the wing generated higher positive drag force and near-zero lift force, which is suitable for landing on the initial landing location without gaining altitude. Through flow field analysis, we show that when a rapidly pitching plate is synchronized with the start of the decelerating motion (i.e., $t^*_{os}$ = 0 case), the plate can transfer a higher added mass energy to the fluid by generating a stronger dipole jet during the initial stage of the unsteady motion. Subsequently, the wing recovers this energy as parasitic thrust, which can be utilized by  hunting birds for acceleration after capturing prey. It demonstrates that birds utilize rapid wing pitching during deceleration to accomplish diverse flying objectives, which can be achieved either by recovering added mass energy later in the maneuver or delaying the formation of vortices towards the end of the motion. 

Our experiments revealed that the ground effect can significantly enhance the aerodynamic performance of rapidly pitching plates in deceleration motion. Regarding ground heights $(0.04 < h^* < 1.5)$, our results showed that the ground proximity affects the initial rise in the lift force, which increased as the wing gets closer to the ground. However, the initial drag force remained independent of ground height. The ground effect impacts the parasitic thrust, which increased for all $\Xi$ values when the wing is close to the ground. Although minor disparities exist between the predicted and experimental results, our present analytical model effectively captures the general trend in the evolution of instantaneous $C_L$ on rapidly pitching plates during deceleration. Notably, the model predicts an increase in the $C_L$ as the wing approaches the ground, accurately reflecting the influence of ground proximity and demonstrating strong agreement with the experimental results. 

Overall this study highlights the significance of tuning rapid pitching with deceleration, as observed in perching and hunting birds, which employ strategies such as the recovery of added mass energy for acceleration or the delayed formation of vortices for smooth landing or perching. This improved understanding of the highly unsteady performance of natural flyers contributes to the design of safer and more efficient flying vehicles.

\section*{Acknowledgments}
Acknowledgments

\bibliographystyle{jfm}
\bibliography{jfm-instructions}

\end{document}